\documentclass[%
 reprint,
 amsmath,amssymb,
 aps,
]{revtex4-2}

\usepackage{graphicx}
\usepackage{dcolumn}
\usepackage{bm}

\usepackage{siunitx}
\usepackage{xcolor}

\begin{document}

\preprint{APS/123-QED}

\title{Generation of Josephson vortices in stacked toroidal Bose-Einstein condensates}
\author{Nataliia Bazhan$^{1, 3}$, Anton Svetlichnyi$^{1}$, Dominik Pfeiffer$^{2}$, Daniel Derr$^{2}$, Gerhard Birkl$^{2}$, Alexander Yakimenko$^{1}$}

\affiliation{$^1$Department of Physics, Taras Shevchenko National University of Kyiv, 64/13, Volodymyrska Street, Kyiv 01601, Ukraine}
\affiliation{$^2$ Technische Universit\"at Darmstadt, Institut f\"ur Angewandte Physik, Schlossgartenstr. 7, 64289 Darmstadt, Germany}
\affiliation{$^3$ Physikalisch-Technische Bundesanstalt, Bundesallee 100, D-38116 Braunschweig, Germany}

\date{\today}

\begin{abstract}
Coupled coaxially stacked toroidal condensates
with persistent currents suggest an appealing physical platform for investigation of various phenomena related to interacting  superflows from Josephson’s effects
in the regime of weak interactions to the quantum Kelvin–Helmholtz instability for merging rings.
We suggest experimentally accessible methods to 
prepare states with different topological charges in two coupled coaxial ring-shaped atomic Bose-Einstein condensates. Our results open up the way to direct observation of rotational Josephson vortices in atomic Bose-Einstein condensates. 
\end{abstract}

\maketitle


\section{Introduction}\label{sec:Introduction}
Coupled toroidal Bose-Einstein condensates (BECs) with Josephson vortices (JVs) in Josephson junctions are the perspective elements for creating the atom 'circuits' in atomtronics \cite{doi:10.1116/5.0026178}. 
Coaxially stacked circular BECs have become the topic of a large body of theoretical studies \cite{PhysRevLett.98.050401, PhysRevA.81.025602, Zhang2013, Polo_2016, PhysRevLett.110.215302, Brand2008, PhysRevA.98.053603, Amico2014SuperfluidQS, Aghamalyan2013, Haug_2018, gallemi2015, PhysRevA.96.013620,PhysRevA.103.023316, Nicolau2020}. 
Tunnelling coupling of the coaxial superfluid rings suggests an intriguing possibility for the investigation of the atomic analogues of the long Josephson junction with varying phase along
the junction and  JVs  between the rings. These remarkable topological structures, also known as rotational fluxons,  found a practical application in superconductors, where JVs (magnetic fluxons) are used in quantum information processing systems.
Previous theoretical studies \cite{PhysRevA.71.011601,PhysRevA.98.053603,PhysRevLett.110.215302,PhysRevA.73.013627,Brand2008,2012JPhB...45c5004Q,2018ScPP....4...18B,PhysRevLett.111.105302,PhysRevA.93.033618, sym11101312,OLIINYK2020105113,Oliinyk2019TunnelingOP} have drawn considerable
interest to JVs in BECs. 
Remarkably, JVs, being extensively investigated for decades, have been directly observed experimentally only very recently in superconductors \cite{2015NatPh..11..332R},  in a polariton superfluid \cite{2019NaPho..13..488C}, and have not yet been demonstrated experimentally in BECs. 

The first direct observation of rotational JVs in bosonic junctions now appears to become a realistic possibility. 
The azimuthal
structure of the tunnelling flows in a double-ring BEC system with topological
charges $m_1$, $m_2$ implies formation of $|m_1 - m_2|$ JVs. Significantly, the existence of JVs is certain to superflows with different topological charges no matter whether tunnelling flow is strong or weak. This suggests that close control over the tunnelling regime is not a necessity for experimental observation of the rotational JVs.
As the barrier decreases and the effective coupling between the rings increases, respectively, the JVs accumulate more and more energy, which simplifies their direct observation. In elongated strongly coupled double-ring BECs, JVs are very robust \cite{OLIINYK2020105113}, and they can be readily observed in contrast to superconducting counterparts.


Essential progress toward an experimental realization of the double-ring system has been made
recently by demonstration of the qubit-based two coaxial lattice rings \cite{Amico2014SuperfluidQS,Aghamalyan2013EffectiveDO} and the single lattice ring with three weak links \cite{Aghamalyan2015AnAF}. However, the possibility of practical realization of the interacting superflows in coaxially stacked atomic rings is an important problem that remains to be studied. 
In the present work we address the following issues: (i) How to generate superflows with different winding numbers in coaxially stacked toroidal BECs, and (ii) what is the physical mechanism of the Josephson vortex emerging in the atomic analogue of the long Josephson junction.

The rest of the paper is organized as follows. The model
is introduced in Sec. II. In Sec. III we present the method for generation of the persistent currents in coupled superfluid rings with population imbalance by a stirring laser beam. In Sec. IV
we introduce the method of controllable asymmetric persistent current decay in a double-ring system. In Sec. V
we apply to the double-ring system a method of stochastic persistent current generation during the merging of initially separated fragments. In Sec. VI we compare different methods for the generation of JVs. The paper is
concluded in Sec. VII.


\section{Model}\label{sec:Model}
We base our model on a series of theoretical works \cite{PhysRevA.91.023607, snizhko2016, Yakimenko2015VorticesIA, Oliinyk2020NonlinearDO, Oliinyk2019SymmetryBI, Oliinyk2019TunnelingOP, PhysRevA.91.033607, sym11101312, OLIINYK2020105113, Oliinyk_2019} exploring a $^{23}$Na Bose-Einstein condensate with similar parameters to those used in the experiment \cite{PhysRevLett.110.025302}.

In the mean-field approximation, the evolution of ultracold Bose-Einstein condensates with weak dissipation can be described by the dissipative Gross-Pitaevskii equation \cite{Choi,Prokakis2008JPhB...41t3002P}:
\begin{equation}
 (i-\gamma)\hbar \frac{\partial {\Psi} }{\partial t}= \left(-\frac{\hbar ^{2}}{2M} \nabla ^{2}+V_\textrm{ext}(\mathbf{r})+{g}|{\Psi} |^{2} -\mu\right) {\Psi},  \label{GPE}
\end{equation}
where $\gamma\ll 1$ is a phenomenological dissipation parameter,  ${g} = {4\pi\hbar^2a_s}/{M}$ is the coupling strength, $a_s = \SI{2.75}{\nano\meter}$  is the $s$-wave scattering length of  $^{23}$Na atoms with atomic mass $M$, and $\mu(t)$ is the chemical potential of the whole system at time $t$.  The dissipative parameter $\gamma$
determines the relaxation time of the vortices: the greater
$\gamma$ is, the less time it takes for a vortex to drift from the high-density condensate annulus to the low-density periphery. \textcolor{black}{The  damping coefficient increases with the temperature since it represents the rate at which the excited components turn into the
condensate. A theoretical estimate \cite{Choi} gives $\gamma\approx 0.03$ for temperature $T\sim T_c/10$.}  For the remainder of this article,  we use $\gamma = 0.03$, however, as we verified, our main results do not depend qualitatively on the specific value of $\gamma\ll 1$. The external potential $V_\textrm{ext}(\mathbf{r}, t) = V_t(\mathbf{r})+ V_s(z)$ includes the toroidal trap potential and the potential of narrow blue-detuned light sheet, which splits the condensate into two separate ring-shaped condensates. Thus, the ring BECs are stacked coaxially along the z-direction. Here we use the parameters of the experiment with toroidal BEC reported in  Ref.~\cite{PhysRevLett.110.025302} for the trapping potential:
\begin{equation}\label{Vt}
V_{\textrm{t}}(\textbf{r})=\frac12 M\omega_r^2(r_\perp-R)^2+\frac12 M\omega_z^2 z^2.
\end{equation}
with  $r_\perp=\sqrt{x^2+y^2}$.
The frequencies $(\omega_r,~\omega_z)=2\pi\times (130,~633) \, \si{\hertz}$ correspond to the oscillatory lengths $l_r=\sqrt{\hbar/(M\omega_r)} = \SI{1.84}{\micro\meter}$ and $l_z = \sqrt{\hbar/(M\omega_z)} =\SI{0.83}{\micro\meter}$. The number of atoms is $N = \SI{6e5}{}$, the radius of the ring trap is $R = \SI{19.23}{\micro\meter}$.

In terms of harmonic oscillator units we rewrite $t \to \omega_r t$, $(x,y,z)\to (x, y, z)/l_r$, $V\to V/(\hbar\omega_r)$, $\mu\to \mu/(\hbar\omega_r)$. Thus in dimensionless form (\ref{GPE}) can be written as:
\begin{equation}\label{GPE_dim}
(i-\gamma)\frac{\partial \psi}{\partial t} = \left[-\frac{1}{2} \nabla^2 + V_{ext}(\textbf{r}, t) +  g|\psi|^2 -\mu(t)\right]\psi,
\end{equation}
where { $\psi = l_r^{3/2}  \Psi$}, $V_{ext}(\textbf{r}, t) = V_{t}(\textbf{r}) + V_{s}(z)$. In dimensionless form:
\begin{equation}
V_t(\textbf{r})=(r_\perp-R)^2/2+\kappa^2z^2/2,
\end{equation}
the parameters of the stationary toroidal potential are the dimensionless radius
$R=10.45$ and the ratio of oscillator frequencies $\kappa=\omega_z/\omega_r=4.88$. Here and further in the text we write all dimensionless lengths in units of $l_r$ and all dimensionless potential amplitudes in units of $\hbar\omega_r$. 

Note that adding a 1D optical lattice in the z-direction to a prolate (elongated in a z-direction) toroidal trap would give a possibility to create a stack of separated coaxial rings. In the present work, we concentrate on the simplest case of two coupled rings. The potential of the blue-detuned light sheet splitting the BEC into two coaxial rings is as follows:
\begin{equation}\label{Vs}
V_s(z) = U_s e^{-\frac{1}{2}{(z-z_{sh})^2}/{a^2}},
\end{equation}
where  $U_s = 80$ is the dimensionless barrier amplitude. The coupling strength decreases when the barrier amplitude and/or barrier effective width increase. In the present work we change the ring-ring interaction strength from strong to weak coupling by tuning the effective barrier width, $a$ in the experimentally accessible range (e.g. in units of $l_r$, $a=0.19$ corresponds to the barrier parameters reported in Ref. \cite{2007Natur.449..579L}).  The population imbalance in rings depends on the parameter $z_{sh}$ -- the $z$-coordinate of the peak intensity of the barrier. A symmetric system with equal populations in both rings corresponds to $z_{sh} = 0$.

It is worth mentioning, that we do not include the effects of gravity in the model. Thus the plane of each ring is only given by its $z$-coordinate and we refer to the 'upper' and 'lower' ring only with respect to their position in the 3D figures of this article. 



\section{Stirring of an asymmetrically populated double-ring system}

\begin{figure*}
	\includegraphics[width=2\columnwidth]{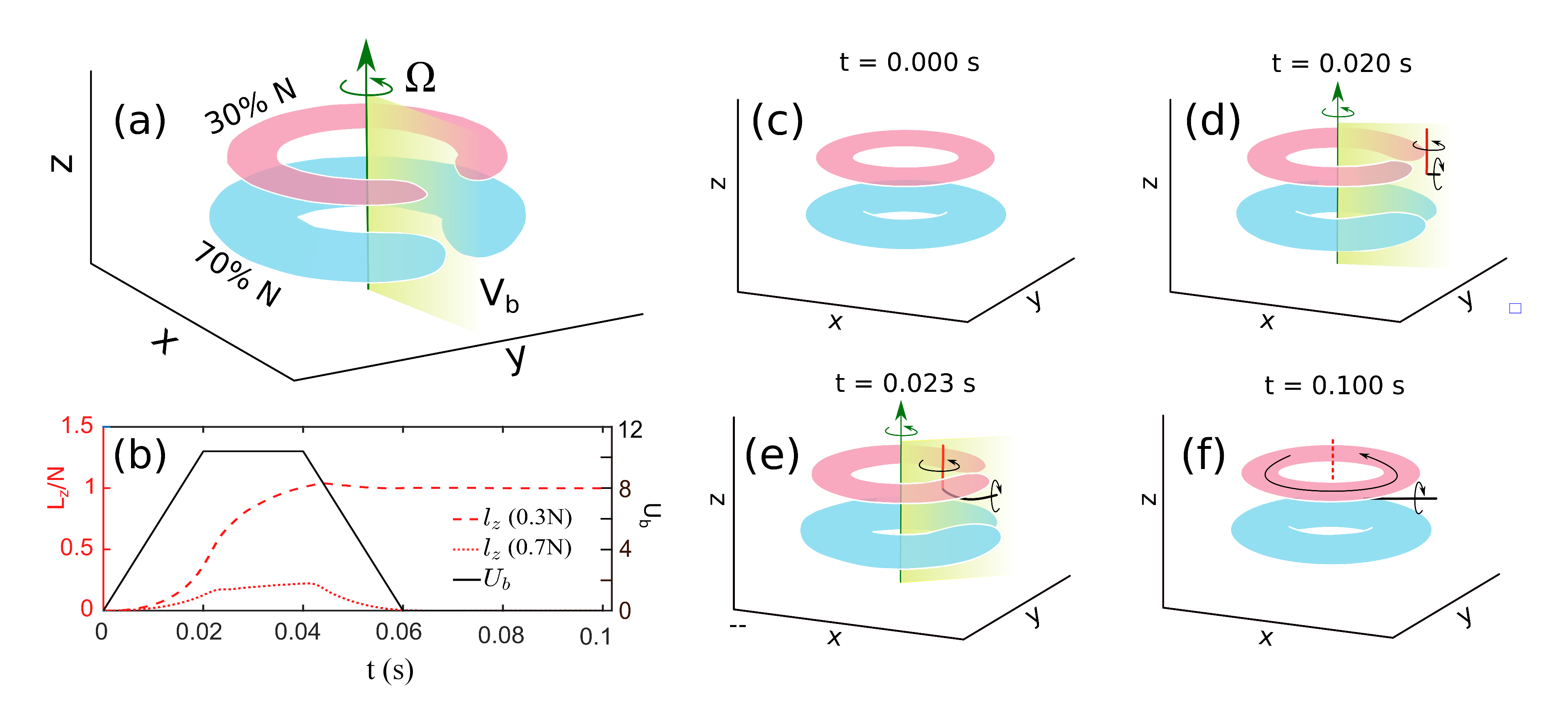}
	\caption{ 
     \textcolor{black}{ (a) Schematic illustration of generating the persistent current by the stirring potential $V_b$ with angular velocity $\Omega/(2\pi) =  2$ Hz in a system of coaxially stacked toroidal condensates divided by a splitting potential $V_s$ with $a = 0.30$ with the initial (0, 0) state. The upper (pink) and lower (blue) rings consist of $30\%$ and $70\%$ of atoms, respectively. Isosurfaces are plotted for a value of $|\Psi|^2$ equal to $44\%$ of the peak density. The stirring potential $V_b$, depicted by the yellow area and green arrows, rotates anti-clockwise with constant angular velocity $\Omega$. (b) Evolution of the angular momentum per particle of the rings (dotted red line for the lower ring, dashed red line for the upper ring) and the stirring potential amplitude $U_b(t)$ (solid black line) in dimensionless units.  Snapshots (c-f) schematically show the persistent current generation: (c) at $t = 0$ s the system is in the initial (0, 0)-state; (d-e) during stirring the $L$-shaped vortex line gets inside the system through the weak link. Red (black) lines depict vortex cores found by phase unwrapping in horizontal (vertical) planes. Black arrows around the vortex lines depict the direction of the superflow circulation; (f)  after switching off the stirring potential the lower-populated ring acquires a topological charge of $m = 1$ while the higher populated one remains in the zero-vorticity state. 
     The red dashed line illustrates the vertical vortex corresponding to the persistent current in the upper ring (see Fig. \ref{fig:phase_stirring}). Note,  that because of the low density we cannot directly detect the precise position of the central vortex core.
     }
	\label{fig:isosurf}}
\end{figure*}
	

\begin{figure}
	\includegraphics[width=1\columnwidth]{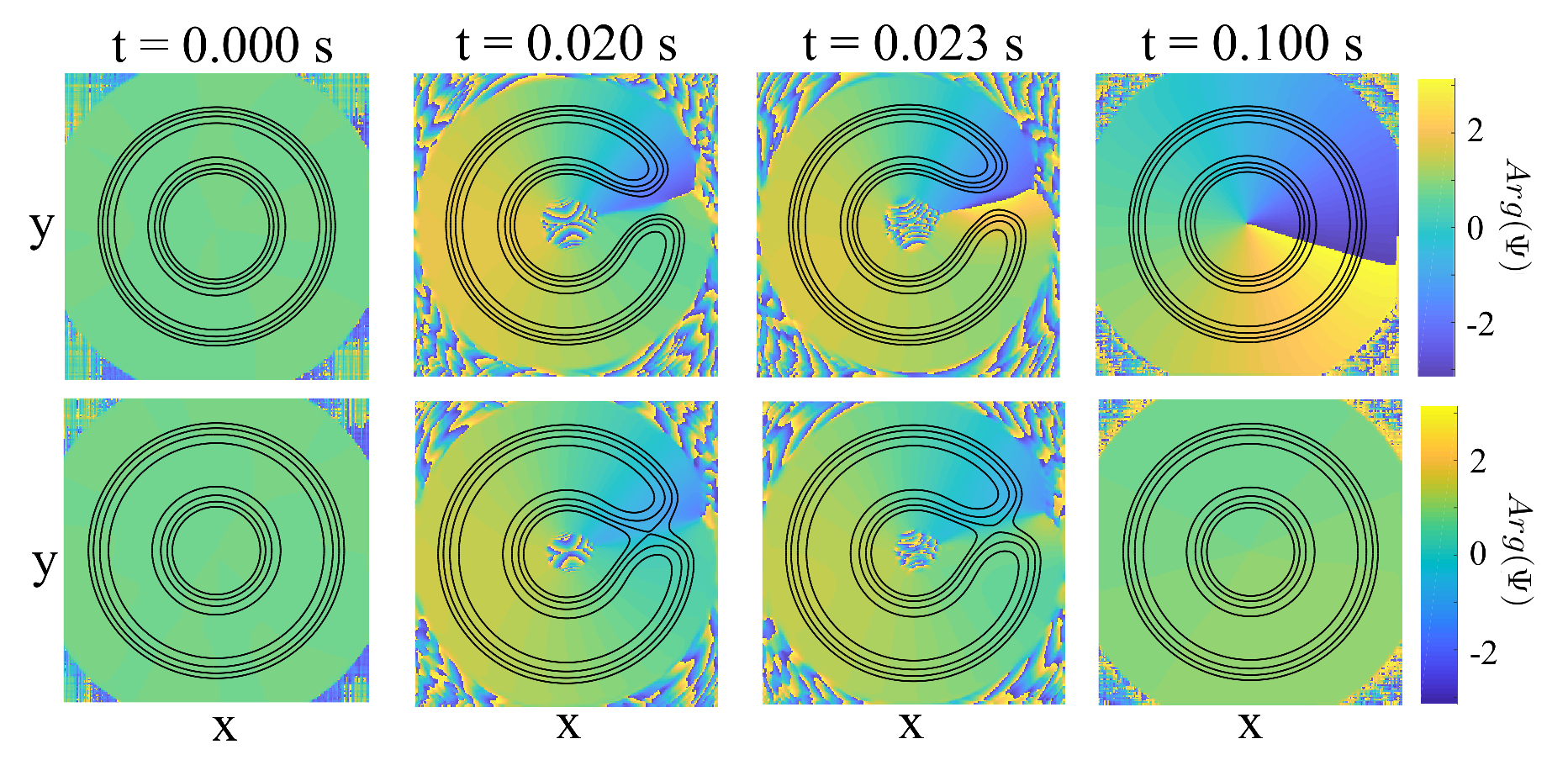}
	\caption{ 
	\textcolor{black}{Snapshots of the color-coded phase and contours of the density distribution in $(x,y)$ plane illustrating dynamics from Fig. \ref{fig:isosurf}. \textcolor{black}{The top (bottom) row correspond to the upper (lower) ring in Fig. \ref{fig:isosurf}. The times match to Fig. \ref{fig:isosurf} (c) to (f)}.}
	\label{fig:phase_stirring}}
\end{figure}

\begin{figure*}
	\includegraphics[width=2\columnwidth]{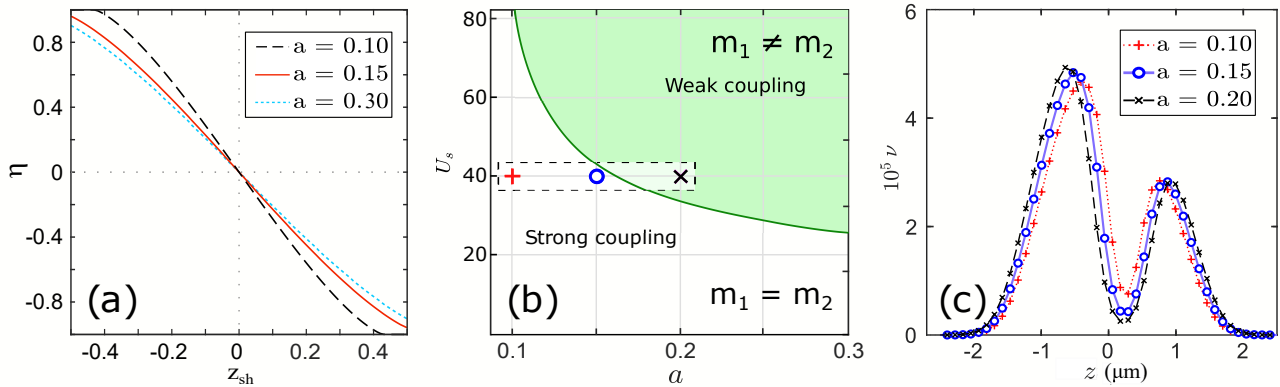}
	\caption{(a) The population imbalance $\eta$ depending on the $z$-coordinate $z_{sh}$ (measured in $l_r$ units) of the splitting light sheet $V_s$ for $U_s = 80$ for different effective widths $a$. (b) The threshold of the JV generation as the function of the splitting barrier amplitude $U_s$ and barrier effective width $a$ after stirring them with angular velocity $ \Omega = 2 \pi \cdot \SI{2}{\hertz}$. The green shaded region corresponds to the final states with different topological charges (weak coupling region), and the white region corresponds to final states with equal topological charges (strong coupling region).
	Parameters $a$ and $U_s$ are measured in $l_r$ and $\hbar\omega_r$ units, respectively. (c) Integrated condensate density distribution $\nu(z)$ along $z$-axis  at fixed barrier amplitude $U_s= 80$ and different values of the barrier width $a$ (the points are indicated in (b) by corresponding symbols).}
	\label{fig:den_critical}
\end{figure*}

\textcolor{black}{
As known, a persistent atomic flow in a toroidal trap can be created by stirring with a repulsive barrier rotating around the ring  \cite{PhysRevLett.110.025302,2014Natur.506..200E}. The presence of a stirring potential produces a localized region of reduced superfluid density with increased condensate velocity in the
condensate annulus -- a rotating weak link.
Tuning the barrier height $U_b$ and angular velocity $\Omega$ one can drive the phase slip corresponding to the topological charge transformation $0 \to 1$ \cite{PhysRevA.91.033607}.
}

\textcolor{black}{
Since the stirring potential produces a deeper weak link in a lower-populated ring, the threshold values of the barrier height $U_b$ and angular velocity $\Omega$ \textcolor{black}{for inducing the phaseslip} decrease with the number of atoms in the condensate.
Thus, using a single laser beam to simultaneously stir two stacked rings with different populations makes it possible to excite the superflow only in the less populated ring, keeping the more populated one in the zero-vorticity state, as illustrated in Fig. \ref{fig:isosurf} and Fig. \ref{fig:phase_stirring}. 
}


\textcolor{black}{
 Let us consider the time-dependent external potential:
$
     V_\textrm{ext}(\textbf{r}, t) = V_{t}(\textbf{r}) + V_{s}(z) + V_{b}(\textbf{r}, t),
$
which includes a toroidal trap $V_t(\textbf{r})$ defined in equation (\ref{Vt}), a splitting potential $V_s(\textbf{r})$ (\ref{Vs}) creating population imbalance, and a time-dependent \textcolor{black}{repulsive (i.e. blue-detuned)} stirring potential $V_b(\textbf{r}, t)$.
For simplicity, $V_b$ is considered to be homogeneous in the radial direction across the toroidal condensate:
\begin{equation}\label{Vb}
V_b(\textbf{r}_\perp,t)=U_b(t)\Theta(\textbf{r}_\perp\cdot \textbf{n})e^{-\frac{1}{2c^2}\left[\textbf{r}_\perp\times \textbf{n}\right]^2},
\end{equation} 
$\textbf{r}_\perp$ is the radius vector in the $(x,y)$ plane;
the unit vector $\textbf{n}(t ) = \{\cos(\Omega t), \sin(\Omega t)\}$ points along the azimuth of the barrier maximum, $U_b$ is the barrier height, $c = 1.85$ describes the barrier width,  the Heaviside theta function $\Theta$ assures a semi-infinite radial barrier potential starting at the trap axis. 
}

\textcolor{black}{
In this section we consider the system having a population imbalance $\eta=(N_1-N_2)/(N_1+N_2)=-0.4$ ($30\%$ of all particles are trapped in the upper ring and $70\%$ - in the lower one) which is done by choosing the proper parameters $a, ~z_{sh}$ of the splitting potential $V_s$ (\ref{Vs}). Figure  \ref{fig:den_critical}(a) shows the population imbalance $\eta$  as a function of $z_sh$ for different widths $a$ and a fixed amplitude $U_s = 80$.}

\textcolor{black}{
As illustrated in Fig. \ref{fig:isosurf}(b) we linearly increase the amplitude of the weak link up to $U_{b, max} \approx 11$ 
during the first $t_{\uparrow}  = 0.02 \, s$. Then $U_b$ remains maximum until the $t_{\rightarrow}  = 0.04 \, s$ and then linearly decreases until $t_{\downarrow}  = 0.06 \, s$. We analyze the system at $t = 0.3\,s$ when its phase distribution stabilizes:
\begin{equation}\label{protocol}
U_b(t) = 
\begin{cases} 
U_{b, max} \cdot \frac{t}{t_{\uparrow}} &\mbox{if } t \in [0, t_{\uparrow}), \\
U_{b, max} & \mbox{if } t \in [t_{\uparrow}, t_{\rightarrow}),\\
U_{b, max} \cdot \frac{t - t_{\rightarrow}}{t_{\downarrow} - t_{\rightarrow}} &\mbox{if }  t \in [t_{\rightarrow}, t_{\downarrow}).\\
\end{cases} 
\end{equation}
}

\textcolor{black}{
Here and thereafter, the numerically detected vortex cores are drawn with solid lines (Fig. \ref{fig:isosurf}(d-f)) and the 'central' vortices are depicted by dashed lines (Fig. \ref{fig:isosurf}(f)). We call the 'central' vortices those vortices which are not detectable directly by phase unwrapping in the low-density regions, but, because of the nonzero phase slip along the ring, these vortices are located inside the toroidal hole.
}

\textcolor{black}{
In Figs. \ref{fig:isosurf}(d-f) we mark the direction of the flow around the vortex lines using black arrows. In Fig. \ref{fig:isosurf} we draw the arrow along the upper ring to emphasize that the dashed 'central' vortex corresponds to a phase slip calculated along the ring, as it can be seen from Fig. \ref{fig:phase_stirring}. Note, that usually vertical vortices and JVs are the parts of a single $L$-shaped vortex core which gets inside the system through the rotating weak link, as is seen in Fig. \ref{fig:isosurf} (c). This example illustrates a general mechanism for the formation of a JV: it appears in the horizontal barrier region between the rings due to the deformation of the same vertical vortex line, which finally forms the persistent current in the toroidal condensate.}

\textcolor{black}{We note that the generation of the JVs in the double-ring system requires a weak-coupling regime with a wide barrier. We are not able to directly follow the streamlines of the tunneling flow \textcolor{black}{with its low condensate density} in the barrier region for the parameters of the simulation in Fig. \ref{fig:isosurf}. However, the direction of the vortex flow induced by JVs can be readily found from the phase distribution, which is shown by the arrows in Fig. \ref{fig:isosurf} (d)-(f). The direct numerical analysis of the tunneling flow in the double-ring system has been performed in Ref. \cite{Oliinyk_2019}, related to the Josephson effect in double-ring system.
}

\begin{figure*}
	\includegraphics[width=2\columnwidth]{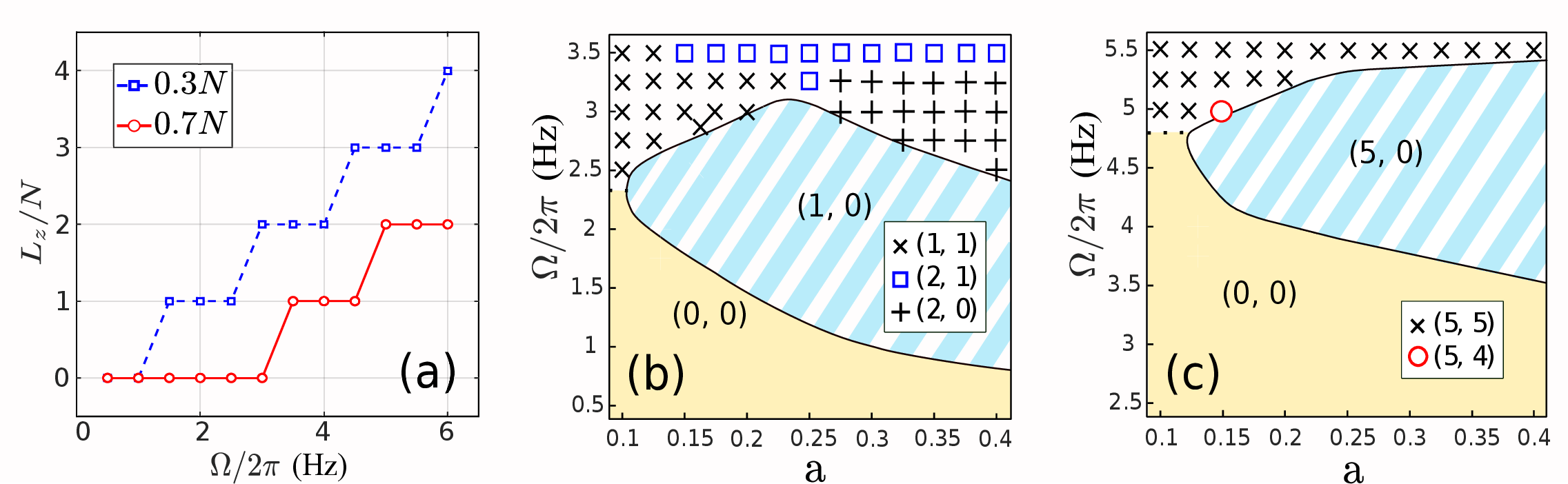}
	\caption{The angular momentum states of the double-ring system at time $ t = 0.3 \, s $ after being stirred by the rotating barrier with angular velocity $ \Omega $. (a) Angular momentum per atom for fixed horizontal barrier width, $ a = 0.30 $. The states are obtained by stirring by a rotating potential with angular velocity $\Omega$ of the double ring system separated by the horizontal barrier of width $a$.  (b) and (c) show the region of occurrence of vortex charges $(m_1, m_2)$ as a function of $\Omega$ and the barrier width $a$ for one rotating barrier and five rotating barriers, respectively. The winding numbers of the rings are indicated in the legend.}
	\label{fig:lzw_pt1}
\end{figure*}
We can create a range of different states by changing the angular velocity $\Omega$ of the stirring potential (Fig. \ref{fig:lzw_pt1}(a)). If the coupling between the rings is strong enough it is impossible to prepare the state with different topological charges using the stirring potential. We analyze different parameters of the splitting potential $U_s$ and $a$ for which the rotating barrier with $\Omega = 2\pi \cdot 2$ Hz can no longer switch the system from state (0, 0) to state (1, 0) (Fig. \ref{fig:den_critical}(b)). A strong coupling regime corresponds to a weakly suppressed condensate density between the rings, as illustrated in Fig. \ref{fig:den_critical}(c), which shows the integrated density distribution along $z$-axis:  $\nu(z) = \int |\psi(x, y, z)|^2 dx dy$. 

For  $a \in [0.1, 0.4]$, $\Omega \in [0.5, 6.0]$ Hz  we find the range of possible combinations of $a$ and $\Omega$ for which stirring the system (0, 0) by the basic protocol transfers it into the (1, 0) state (see Fig \ref{fig:lzw_pt1}(b)). The weaker the connection between the rings, the easier it is to untwist them.

We note that using a modification of this method, as described below, it is possible to transfer the system of two coaxial BEC-rings from $(m_1,m_2) = (0, 0)$ state to $(n, 0)$ state (the topological charge of the upper (lighter) ring is equal to $m_1=n\ge 1$ and the charge of the lower (heavier) ring is $m_2=0$) using \textcolor{black}{multiple stirring potential $V_b^n$ } with $n$ similar equidistant barriers ($V_b^1 \equiv V_b$):
\begin{equation}\label{Vn}
V_b^n(\textbf{r}_\perp,t)=U_b(t)\sum_{i=1}^{n}\Theta(\textbf{r}_\perp\cdot \textbf{n}_i)e^{-\frac{1}{2c^2}\left[\textbf{r}_\perp\times \textbf{n}_i\right]^2},
\end{equation}
where $\textbf{n}_i(t)=\left\{\cos(\Omega t + \frac{2\pi}{n} (n-i) ),\sin(\Omega t + \frac{2\pi}{n} (n-i))\right\}$ and the amplitude of the barriers $U_b(t)$ changes accordingly to the protocol process.

\begin{figure}
	\includegraphics[width=\columnwidth]{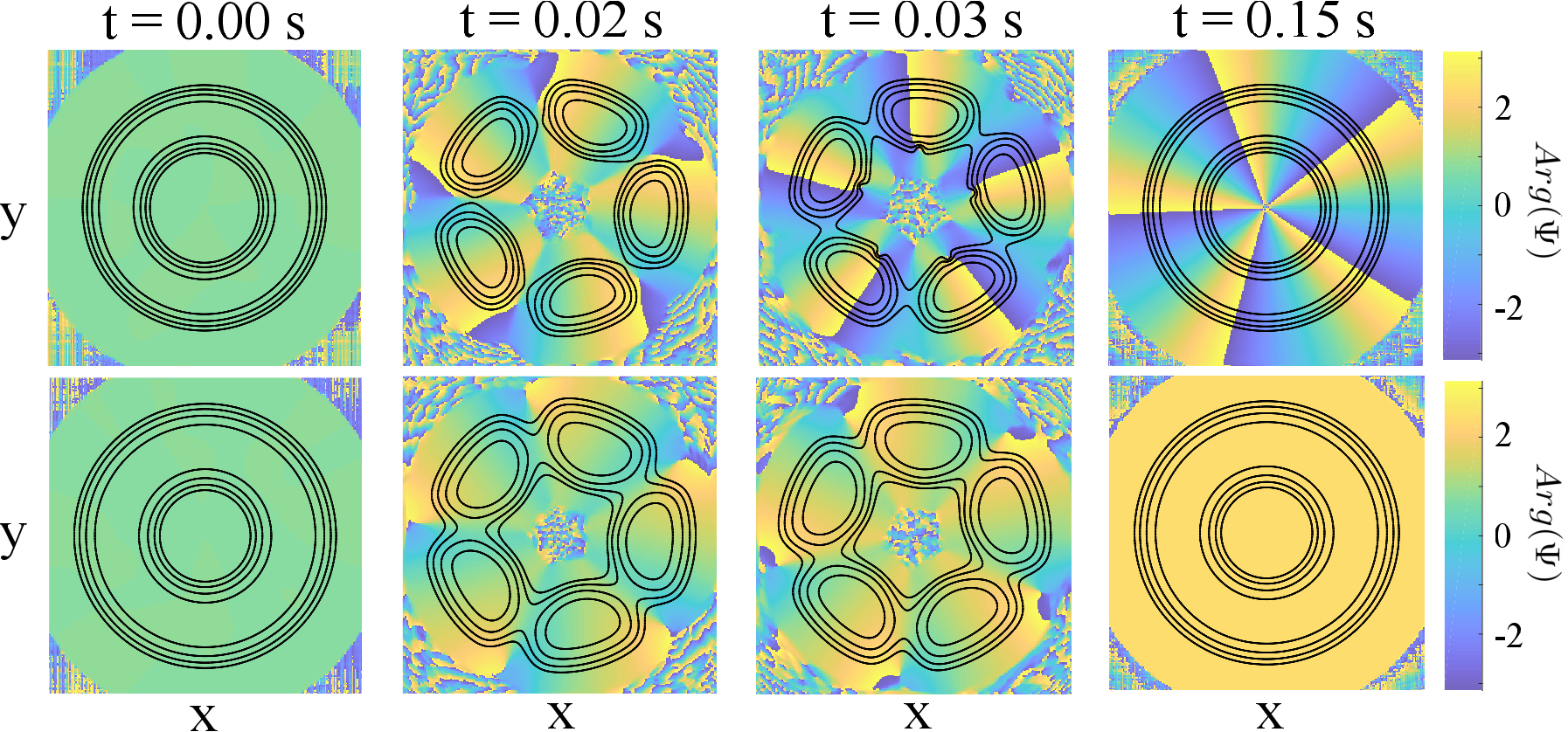}
	\caption{Snapshots of the color-coded phase and contours of the density distribution in $(x,y)$ plane for the  rings during a $(0,0)\to (0,5)$ transition.
	The rotating repulsive potential produces five weak links ($a = 0.30$,  $\Omega = 2\pi \cdot 4.50$ Hz).  Note that five vortex cores enter the ring passing through the weak links in the upper (lighter) ring at $t = 0.03 \, s$, while the vortices fail to get inside the lower (heavier) ring.}
	\label{fig:fluxes5}
\end{figure}

During the stirring $n$ vortices simultaneously find their ways into the central toroidal hole of the less populated ring through $n$ weak links. For example, for $n = 5$  using the proper combinations of $a$ and $\Omega$ (see Fig \ref{fig:lzw_pt1}(c)) it is possible to create the (5, 0) state as illustrated in Fig. \ref{fig:fluxes5}.

\section{Asymmetric decay of persistent current in a double-ring system}
\begin{figure}
    \centering
    \includegraphics[width=\columnwidth]{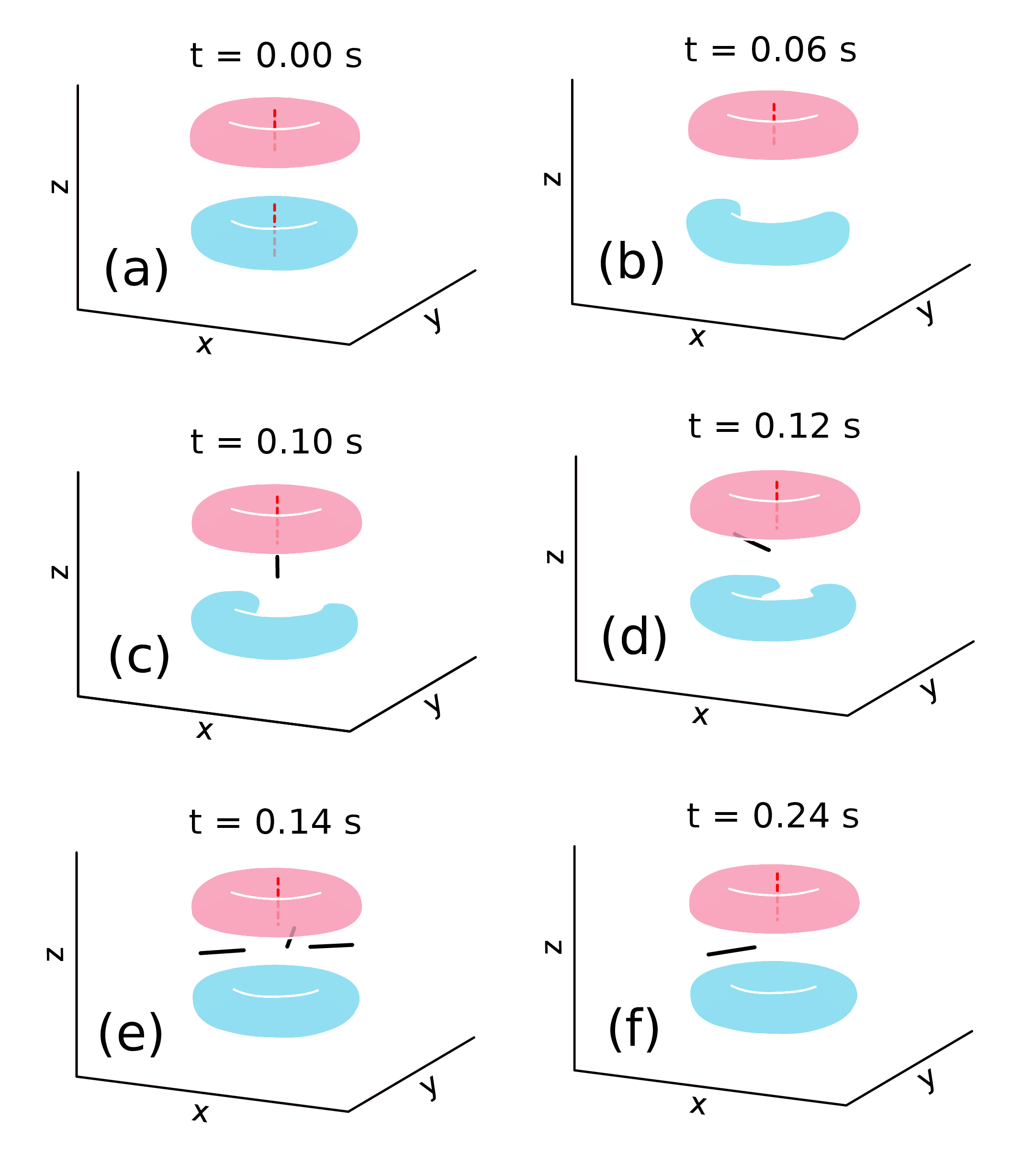}
    \caption{Schematic illustration of the asymmetric persistent current decay in a double ring system. (a) Initial state with single-charged persistent currents in coaxial rings. (b-d) Vortex leaves the lower ring with simultaneous formation of the JV between the rings (black line). (e) When closing the barrier, vortex-antivortex pairs can get between the rings and annihilate right after. (f) Final $(1,0)$ state of the coupled toroidal BEC}
    \label{fig:oneizos}
\end{figure}

In the system with an equal population of atoms in the rings, $z_{sh} = 0$, it is possible to use an alternative procedure for creating coupled persistent currents. Firstly, we prepare a single toroidal condensate with a winding number $m$ using one of the well-established techniques. Then the condensate is separated into two equally populated ($\eta = 0$) coupled rings with initially equal topological charges $m_1=m_2 = m$ by the blue-detuned light sheet (\ref{Vs}), $a = 0.30$. \textcolor{black}{
After preparation, we add a repulsive blue-detuned laser beam $V_{c}(y, z, t)$ to the external potential \textcolor{black}{giving} $V_{ext}(\textbf{r}, t) = V_{t}(\textbf{r}) + V_{s}(z) + V_{c}(y, z, t)$, which \textcolor{black}{before has consisted} of a toroidal trap $V_t$ (\ref{Vt}) and a symmetrical splitting potential $V_s$ (\ref{Vs}).  The repulsive symmetry-breaking potential $V_c$ \textcolor{black}{is offset by $z_c < 0$ and} deforms half of the condensate in the lower ring, transforming a toroidal geometry into a single-connected atomic cloud as shown in Figs. \ref{fig:oneizos}(b-d). It can be written as:}
\begin{equation}\label{Vcut}
    V_{c}(y, z, t) = U_{c}(t)e^{-\frac{(z - z_{c})^{2}}{2w_{z}^{2}}-\frac{(y -y_{c})^{2}}{2w_{y}^{2}}},
\end{equation}
where $U_{c}(t)$ is the time-dependent amplitude of the potential, $w_z = 0.5$ - is the width of potential along $z$-axis 
$w_y = 5$ - is the width of potential along y-axis, 
 $z_c = -1.8$ is equal to the doubled value of the peak density coordinate of the lower ring and should be calculated separately for different values of parameter $a$, $y_c = 15$. The quantized circulation in a ring corresponds
to an $m$-charged vortex line pinned at the center of the toroidal condensate, where the vortex energy has a local minimum. Since the vortex core is bounded by the toroidal potential barrier, even multicharged ($m > 1$) metastable vortex states
can be very robust. 
The asymmetric transformation of the trapping potential of the lower ring vanishes part of an effective energetic potential barrier allowing vortex lines to leave the deformed subsystem.
When the vortices escape from the deformed subsystem, we switch off the deforming repulsive potential and obtain a double-ring system with asymmetric winding numbers: $m_1\ne m_2$, as illustrated in Fig. \ref{fig:oneizos}.


We require, that the potential should be wide enough to ensure that the toroidal topology of the lower ring is transformed into a single-connected cloud so that all vortices will leave the lower condensate ($m_2=0$ in the final state). 

The amplitude $U_c(t)$ linearly increases and decreases according to the protocol (\ref{protocol}) where $U_{b, max} = 200$ in dimensionless units and $t_{\uparrow} = 0.01 \, s$, $t_{\rightarrow} = 0.03 ~ s$ and $t_{\downarrow} =  0.13 ~ s$. 
The time intervals are chosen such that, on the one hand, all the vortices have time to leave the lower ring before the potential is turned off, and on the other hand, the system has time to relax. If one chooses to turn off the potential too quickly, producing a strong perturbation, extra vortices (or anti-vortices) may enter the lower ring, and the number of JVs will be greater (lesser) than necessary. \\

\begin{figure}
    \centering
    \includegraphics[width=\columnwidth]{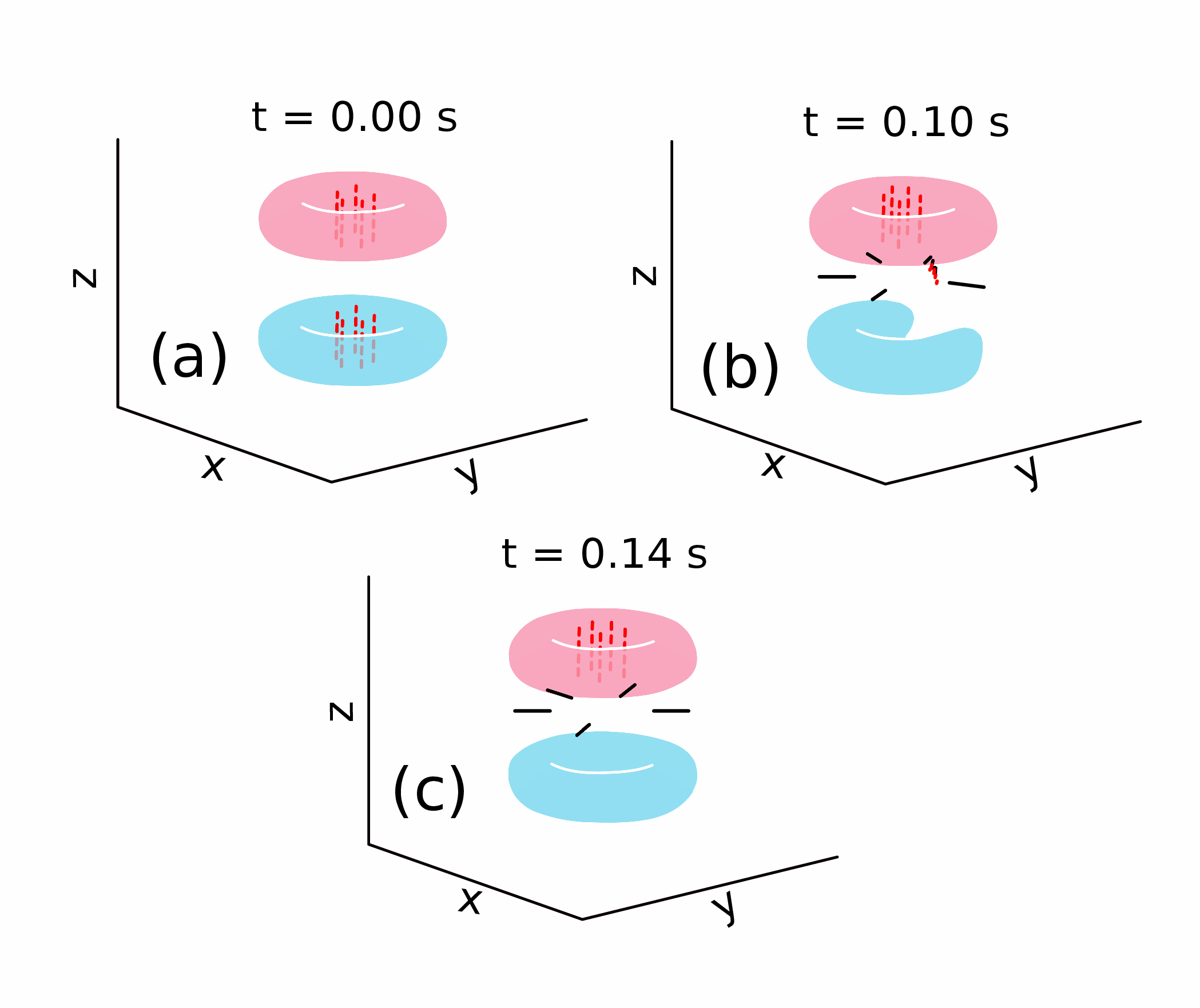}
    \caption{The same as in Fig. \ref{fig:oneizos} for generation  (5,0)-state.}
   \label{fig:fiveiso}
\end{figure}

\begin{figure}
    \centering
    \includegraphics[width=\columnwidth]{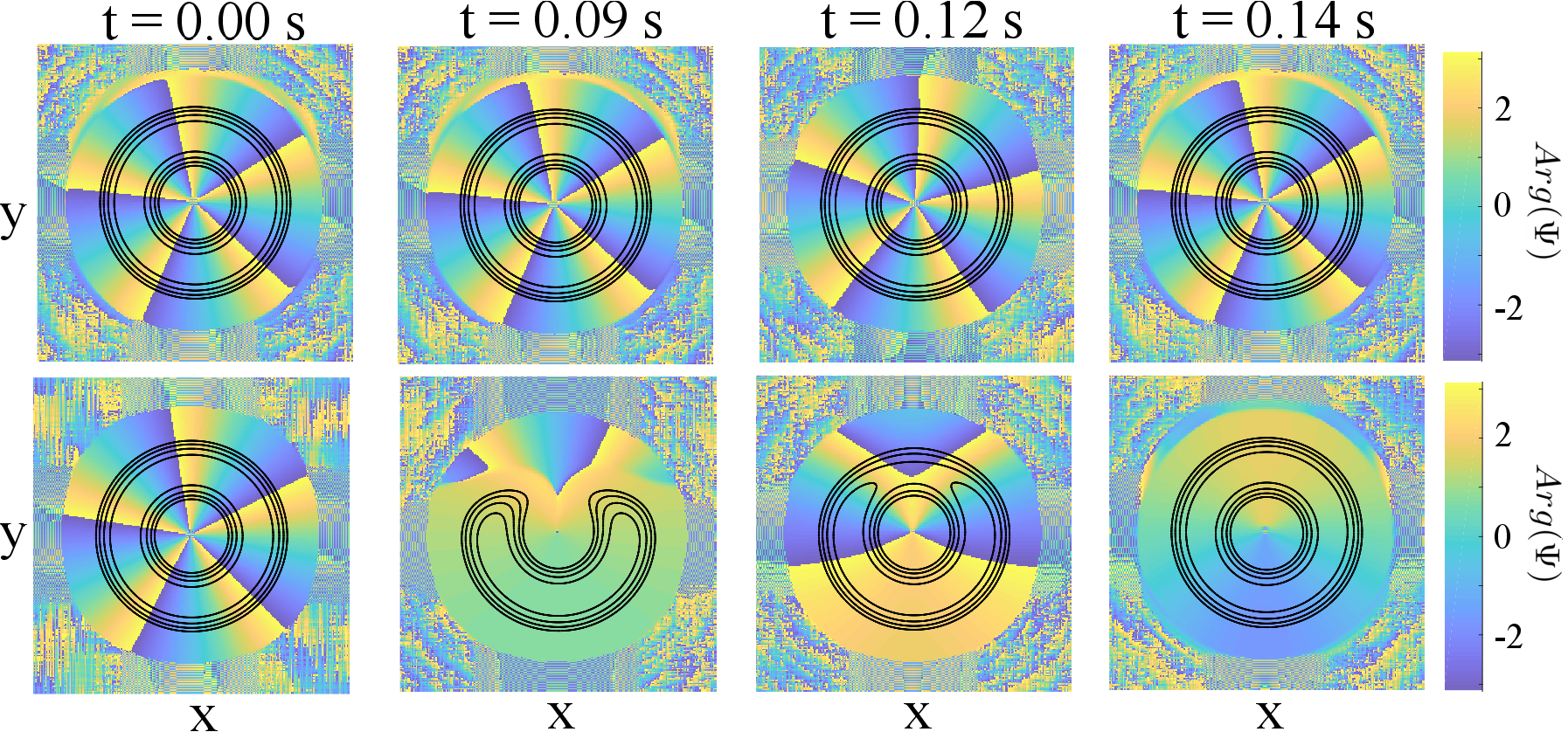}
    \caption{The color-coded phase distribution and density distribution contour lines in cross-sections corresponding to the peak densities for the upper and lower rings illustrating the dynamics shown in Fig. \ref{fig:fiveiso}. The system is transferred from  (5, 5) to (5, 0) state.}
   \label{fig:ph_5}
\end{figure}

To illustrate a typical example of the generation of the double-ring system with a five-charged persistent current in the upper ring and non-rotating lower ring we show snapshots of the density isosurface for different times in Fig. \ref{fig:fiveiso}.  As seen from the phase distributions combined with the condensate density isolines of the rings (Fig. \ref{fig:ph_5}), all the vortices leave the lower ring and the system is transferred from $(5,5)$ to  $(5,0)$ state. In our simulations for the parameters of the condensate used in the present work, we obtained various final states with up to six JVs, corresponding to the $(m_1=6, m_2=0)$ final state.


\section{Stochastic generation of Josephson vortices}
\begin{figure}
    \centering
    \includegraphics[width=\columnwidth]{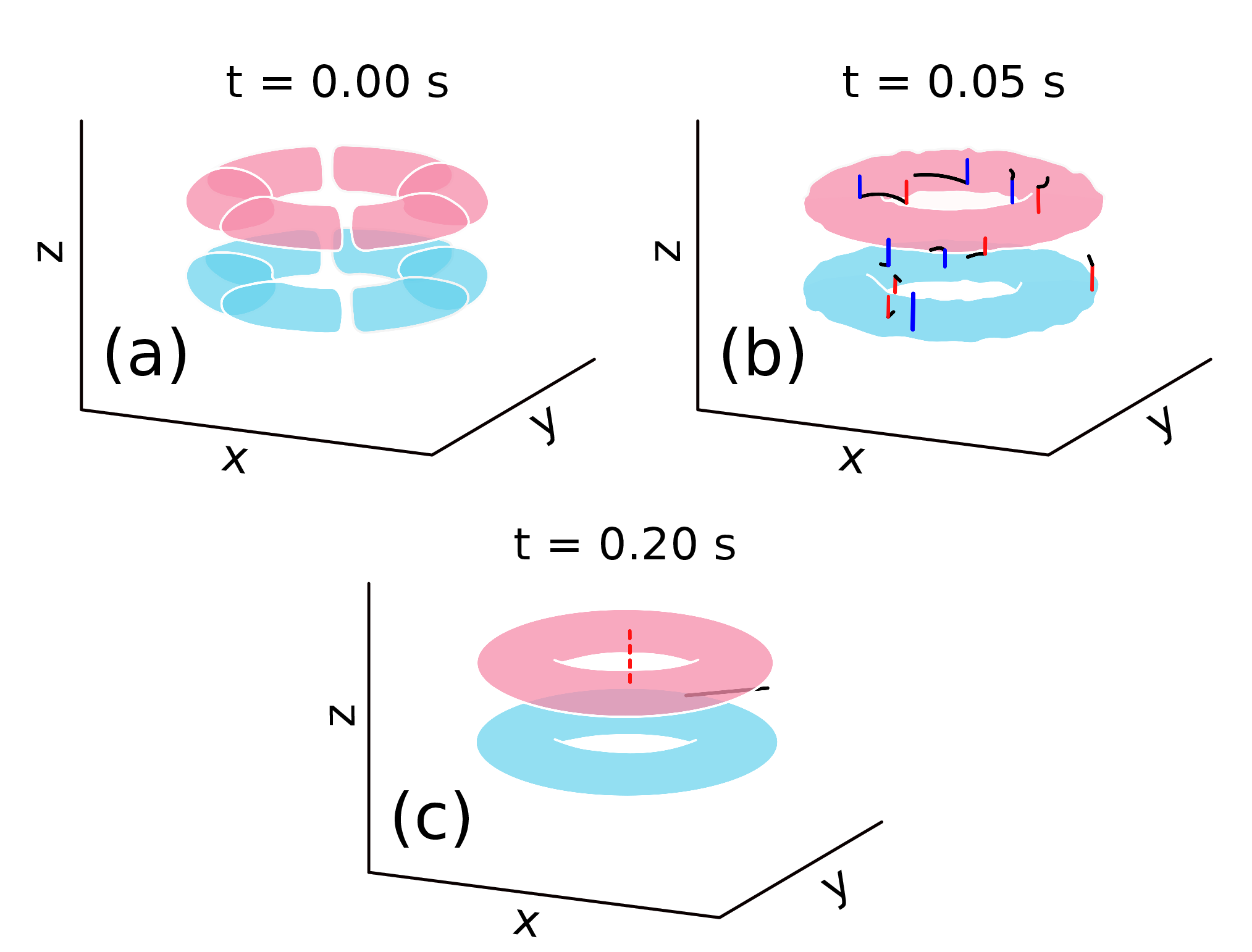}
    \caption{Schematic illustration of stochastic generation of a JV for a weak coupling protocol. \textcolor{black}{Isosurfaces are plotted for $44\%$ of the peak density. Solid lines depict numerically calculated vortices by phase unwrapping. A dashed vertical line inside the ring depicts a nonzero phase slip along it.}  (a) At the time $t = 0 \, s$ the upper and lower rings are split into six fragments by vertical repulsive barriers. Each of these 12 fragments has a random phase in the initial state.  The vertical barriers are switched off rapidly forming the double-ring system.  (b) A large number of vortex excitations are seen in the bulk of the condensate and the low-density region between the rings. (c) Most of the vortices leave the condensate and after the relaxation, the system gains a single-charged vortex line consisting of a vertical vortex (red) in the upper ring and one JV (black) between the rings.}
    \label{fig:stoch_gen}
\end{figure}

In this section, we generalize for the double-ring system a method of stochastic generation of superflows suggested previously \cite{PhysRevLett.119.190403} for a single toroidal condensate. We consider relaxation  of two equally populated ($\eta = 0$) uncorrelated condensates after connecting them in coaxially stacked
ring-shaped condensates.
A typical example of condensate evolution is illustrated in Fig. \ref{fig:stoch_gen}.

First, we prepare the state in the external potential $V_{ext}(\textbf{r}, t) = V_{t}(\textbf{r}) + V_{s}(z) + V_{b}^n(\textbf{r}, t)$. Here the condensate placed \textcolor{black}{in toroidal trap $V_t$ defined in equation (\ref{Vt}) is initially completely separated by both splitting potential $V_s$ (\ref{Vs}) and vertical repulsive multiple potential $V_b^n$ (\ref{Vn}) with 6 non-rotating barriers ($n = 6$, $\Omega = 0$).}
The potential widths $(a = c = 0.185)$ and amplitudes $(U_s = U_b = 530)$ are high enough to split the condensate into 12 fragments without coherent coupling between them. The initial phases are constant per fragment and are randomly selected from a uniform distribution in the range $[0, 2\pi]$.

We consider two protocols with different couplings between the rings. In both protocols the amplitude of the vertical barriers linearly decreases to $0$  from time $t_{\rightarrow} = 0.01 \, s$ to $t_{\downarrow} = 0.04 \, s$. In the first protocol, the splitting barrier remains the same during the merging, thus the coupling between the rings is weak during the experiment. In the second protocol we linearly decrease the width and the amplitude of splitting potential from $(a = 0.185, ~ U_s = 530)$ to $(a = 0.100, ~ U_s = 80)$ while turning off the vertical barriers. In this case, the density between the rings is $\approx 2$ times less than the critical density found for an imbalanced system (see Fig. \ref{fig:den_critical}), thus we consider this protocol describing the system close to the strong coupling. 

\begin{figure}[htb]
    \centering
    \includegraphics[width=1\columnwidth]{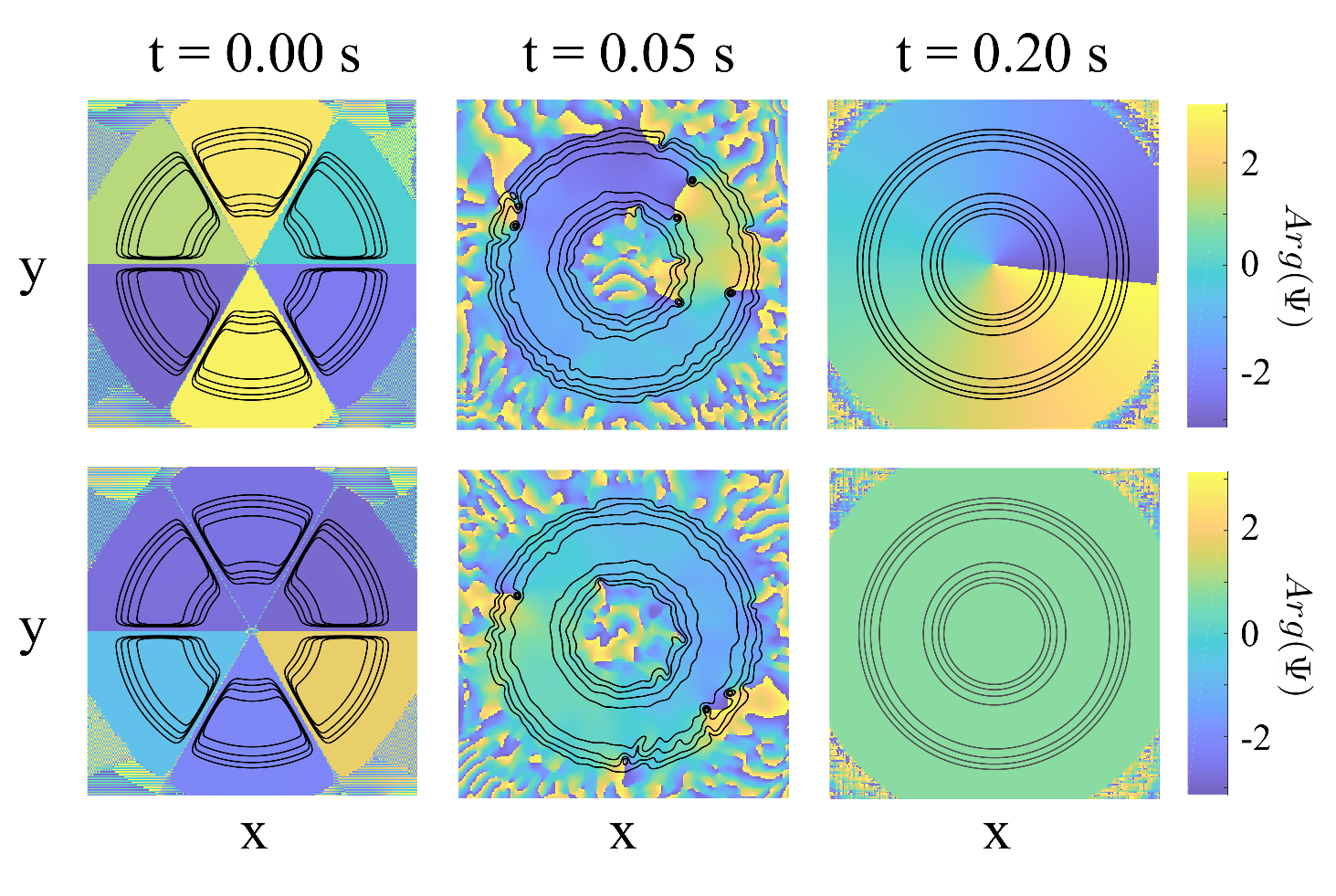}
    \caption{
    The color-coded phase distribution and density distribution contour lines in cross-sections corresponding to the planes of peak densities for the upper and lower rings with similar populations. The system transfer from the (0, 0) state to (1, 0) state due to the merging of separated condensate fragments with random initial phases.
    }
    \label{fig:phase_stoch_gen}
\end{figure}

After the system relaxes (at about $t = 0.2\,s$) we measure the topological charge of both rings.
Rapid switching off of the vertical barriers leads to the excitation of a large number of vortices and antivortices in the double-ring system (Fig. \ref{fig:stoch_gen} (b)). As an example for generating the final state $(1, 0)$, in Fig.  \ref{fig:stoch_gen}, after relaxation, there is a vortex line in the upper ring while the lower ring appears in the ground state. The same dynamic is demonstrated in Fig. \ref{fig:phase_stoch_gen} in phase representation.



\begin{figure}
	\includegraphics[width=\columnwidth]{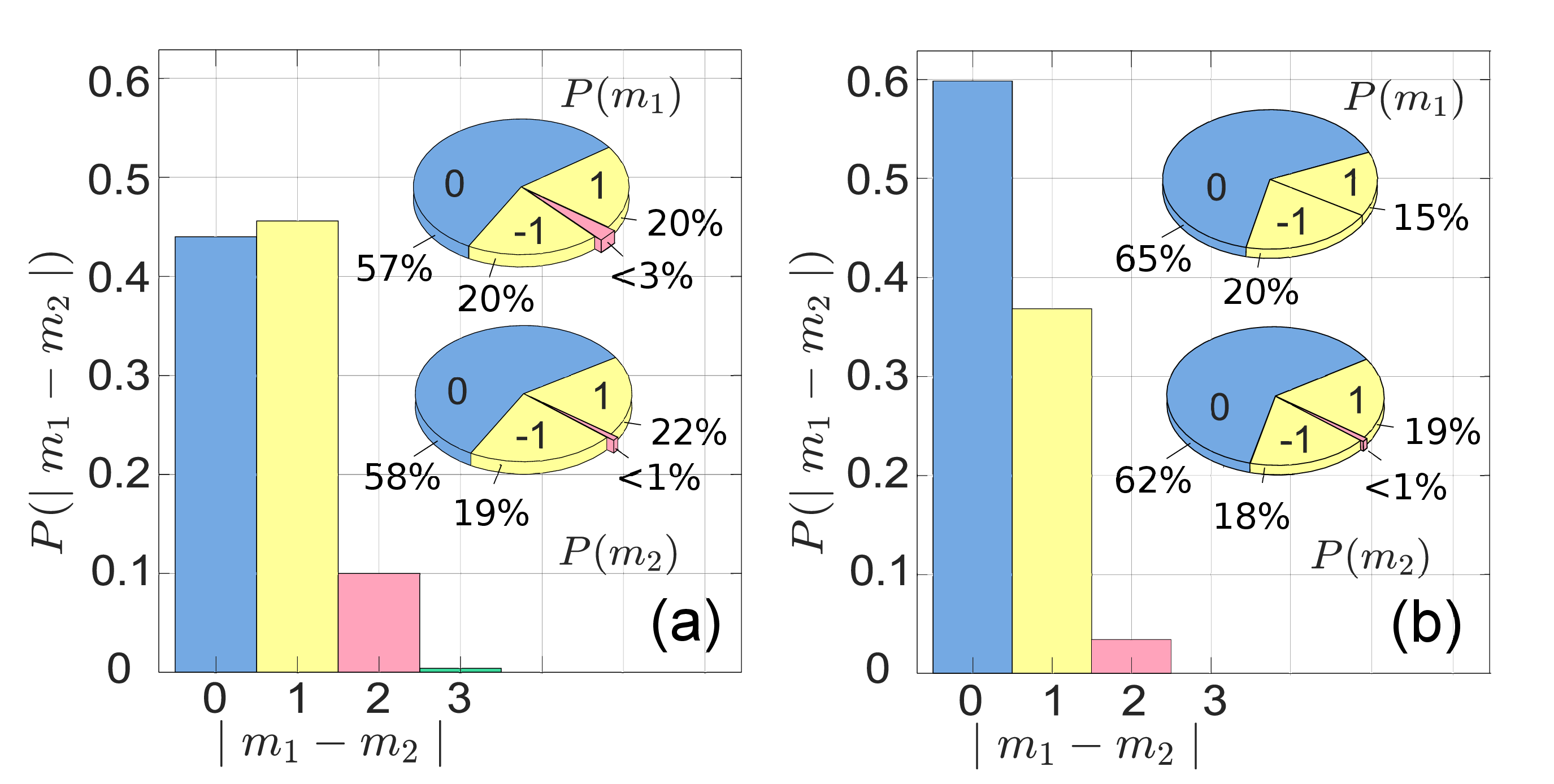}
	\caption{ The histograms show the probability of getting the final state $P(|m_1-m_2|)$ with a specific number of JVs $|m_1-m_2|$. The insets show the probabilities of stochastic generation of the persistent currents in the upper and lower ring $P(m_1)$ and $P(m_2)$ with topological charges $m_1$ and $m_2$ respectively. The results are obtained for a sample of 500 simulations for two different protocols (a) the splitting barrier width and amplitude are fixed at $a= 0.185$ and $U_s=530$ (weak coupling); (b) the splitting barrier width and amplitude $(a, U_s)$ simultaneously decrease from the $(0.185, 530)$ to $(0.100, 80)$ during switching off the vertical barriers (strong coupling). }
   
    \label{fig:stat}
\end{figure}

We have performed two series of $500$ 
simulations for each of the two protocols (weak and strong coupling) with different initial phases of the fragments and calculated the probabilities of getting topological charges $m_1, ~m_2$ in each ring and the probabilities of getting $\mid m_1 - m_2 \mid$ JVs (see Fig. \ref{fig:stat}). For the first protocol, the splitting barrier width and the amplitude $(a, U_s)$ are fixed at $(0.185, 530)$ respectively (weak coupling). For the second protocol, the splitting barrier width and amplitude $(a, U_s)$ simultaneously decrease from the $(0.185, 530)$ to $(0.100, 80)$ during switching off the vertical barriers (strong coupling). The results show high probabilities for a final zero-vorticity state or single-charged persistent currents in one of the rings ($m_i=0$ appears with about 60 $\%$ , $m_i=+ 1$, and $m_i = -1$ with about 20$\%$ for both protocols), and very low probability of formation of multi-charged ($|{m}_i|>1$) persistent currents. That means that the final states with one JV mainly consist of the states where the one ring has $m_i = 0$ topological charge and the other has $m_i = \pm 1$. In the case of the weak coupling protocol (Fig. \ref{fig:stat}(a)) the final state with one JV emerges with high probability ($45\% \pm 5\%$). For the second protocol  (Fig. \ref{fig:stat}(b)) the coupling between the rings is stronger, thus the probability of getting similar topological charges in both rings increases. This results in a lower probability of getting the final state with one JV ($35\% \pm 10\%$). 

\section{Comparison of the JV generation methods}
Let us summarize and compare three methods for the preparation of the coupled superflows with different topological charges suggested in the present work. We note that different angular momentum states could be generated while the rings are more separated (so that the splitting potential is much wider than the healing length in the condensate $a\gg \xi$), then brought together by decreasing barrier width. 
In the present work, we investigate the coupled rings with considerable tunneling flow in the barrier region, which reveals a generic physical mechanism of dynamical formation of JV  in an atomic long Josephson junction. It is remarkable that both, the horizontal JV and vertical vortex line associated with persistent current, simultaneously emerge as the result of bending and further transformation of the single vortex line  in the toroidal condensate, as illustrated in Fig. \ref{fig:isosurf} (b)-(e).

First, we study a persistent current generation by a stirring potential in a  double-ring system with population imbalance. The asymmetry of the density distribution in the top and bottom rings makes it possible to excite the vorticity by applying a stirring laser beam. We demonstrate a possibility for generating a vortex in the lower-populated ring only, keeping the higher-populated subsystem in the zero-vorticity state. This method allows creating a controlled number of JVs in the system with zero angular momentum initial state by tuning the angular velocity and amplitude of the stirring potential.

The second method is based on asymmetric persistent current decay in a double-ring system with initially equal nonzero topological charges. 
Using a repulsive blue-detuned laser beam, we transform one of the toroidal atomic clouds into a single connected condensate, which drives a persistent current decay in the deformed ring, keeping an unchanged angular momentum state in the other ring.  

The third method is based on the stochastic generation of the persistent current as the result of merging initially separated fragments with random phase differences. Rapid switching off of the vertical barriers leads to the formation of the double-ring system with a large number of vortices and antivortices. Dissipative effects cause a drift of these vortex excitations from the high-density condensate annulus to the low-density periphery. It turns out that, if the coupling between the rings remains weak,  the final state exhibits a single-charge persistent current in one ring and a non-rotating condensate in the other ring with a probability of about $45\% \pm 5\%$. For the system with stronger coupling, this probability is lower. 

All these methods can be realized in present state-of-the-art experimental setups and the choice between the methods can be made by taking into account accessible experimental techniques and specific features of the physical problem under consideration. The first method is based on the asymmetry of the density distributions in the coupled rings and allows the generation of different angular momentum states provided the coupling strength between the rings is sufficiently weak. Since the second method is based on the asymmetric decay of the persistent current in one of the rings it can be used for creating the non-rotating states in one ring coupled with the higher-charged persistent current in another ring. Finally, the stochastic generation of the vortices after merging the initially uncorrelated condensates suggests a simple way for the generation of the state with one or two JVs.

\section{Conclusions}
We have investigated the generation of the Josephson vortices in a system of coupled coaxial toroidal BECs. Our analysis reveals a generic physical mechanism of dynamical formation of Josephson vortices  in an atomic long Josephson junction 
and opens a perspective for experimental realization of persistent currents of ultracold atomic gases in a double-ring geometry. This system provides a novel research platform for the investigation of the interacting superflows in a tunable and controllable environment. Using accessible experimental techniques, it is possible to consider a variety of physical phenomena in this setting: from Josephson’s effects
in the regime of weak interactions to quantum Kelvin–Helmholtz instability for merging rings.
Furthermore, the rotational Josephson vortices in the atomic Josephson junction suggest an interesting platform for rotation measurements with atomic matterwaves. 

\section{Acknowledgment}
This work was
supported by the National Research Foundation of Ukraine (2020.02/0032) and the Deutscher Akademischer Austauschdienst
(DAAD), Germany. 
A.Y. and N.B. thank the Institut für Angewandte Physik, Technische Universit\"at Darmstadt
 for warm hospitality during their stay in Darmstadt.

\bibliography{Ring_Refs}

\end{document}